\documentclass[amsmath,10pt,twocolumn,superscriptaddress,footinbib,pre]{revtex4-1}
\usepackage{amsmath,amsfonts,bm}
\usepackage{graphicx}
\usepackage{color}
\usepackage{subfigure}

\begin{document}
\title{Preserving Correlations Between Trajectories for Efficient Path Sampling}
\author{Todd R. Gingrich}
\affiliation{Department of Chemistry, University of California, Berkeley, 
California 94720}
\author{Phillip L. Geissler}
\affiliation{Department of Chemistry, University of California, Berkeley, 
California 94720}
\affiliation{Chemical Sciences Division, Lawrence Berkeley National 
Laboratory, Berkeley, California 94720}

\begin{abstract}
Importance sampling of trajectories has proved a uniquely successful
strategy for exploring rare dynamical behaviors of complex systems in
an unbiased way.  Carrying out this sampling, however, requires an
ability to propose changes to dynamical pathways that are substantial,
yet sufficiently modest to obtain reasonable acceptance
rates. Satisfying this requirement becomes very challenging in the
case of long trajectories, due to the characteristic divergences of
chaotic dynamics.
Here we examine schemes for addressing this problem, which engineer
correlation between a trial trajectory and its reference path, for
instance using artificial forces.  
Our analysis is facilitated by a modern perspective on Markov Chain
Monte Carlo sampling, inspired by non-equilibrium statistical
mechanics, which clarifies the types of sampling strategies that can
scale to long trajectories.  Viewed in this light, the most promising
such strategy guides a trial trajectory by manipulating the sequence
of random numbers that advance its stochastic time evolution, as done
in a handful of existing methods. In cases where this ``noise
guidance'' synchronizes trajectories effectively, as the Glauber
dynamics of a two-dimensional Ising model, we show that efficient path
sampling can be achieved for even very long trajectories.
\end{abstract}
\maketitle 

\section{Introduction}

Recent advances in non-equilibrium statistical mechanics have given 
fresh perspectives on computational procedures applied to fluctuating 
molecular systems.
The Jarzynski work relation and the Crooks fluctuation theorem, for 
instance, provide routes to compute equilibrium quantities from 
non-equilibrium measurements~\cite{Jarzynski1997,Liphardt2002,
Collin2005,Park2003,Oberhofer2005,Nilmeier2011}.
Here, we demonstrate that traditional Metropolis-Hastings Markov Chain 
Monte Carlo (MCMC) can be similarly viewed as a procedure to extract 
equilibrium sampling from generically non-equilibrium processes.
Monte Carlo trial moves drive a system away from the steady state
distribution, and an entropy production can be assigned to these
driven transformations.  
This interpretation provides an elegant way to understand challenges 
encountered in MCMC sampling, one that is especially revealing for MCMC 
sampling of trajectories.  
Path sampling methods suffer routinely from profound inefficiency 
when trajectories of interest become long.  
From a non-equilibrium perspective on MCMC, we provide simple and 
quantitative ways to understand the inefficiency.

Importance sampling of trajectories has enabled studies of a myriad of 
dynamical processes in physics and chemistry~\cite{Bolhuis2000,
Geissler2001,Allen2005,Allen2008,Hu2008,Hedges2009}.
In particular, reaction rates and mechanisms can be found by transition 
path sampling (TPS), which examines the subensemble of trajectories that 
complete a reaction~\cite{Bolhuis2002}.
The practicality of TPS depends intimately on the design of the Monte 
Carlo (MC) move set.
Namely, the moves must generate correlated trajectories so that a trial 
trajectory is likely to exhibit similar dynamical behavior as the 
previously sampled trajectory.
Chaotic divergence and microscopic reversibility of equilibrium
dynamics informs the construction of two such moves, the so-called
``shooting'' and ``shifting'' moves~\cite{Bolhuis2002}.  
These methods generate correlated trajectories by propagating 
alternative histories from highly correlated initial configurations.  
For sufficiently short trajectories, the imposed correlation at one 
time serves to strongly correlate the trajectories at all times.
Long trajectories, however, are problematic: trial trajectories either 
lose all useful correlation with the reference path, or else they coincide
so closely with the reference that changes are impractically 
small~\cite{Grunwald2008}.  
In both cases the efficiencies of shooting and shifting moves plummet 
as trajectories grow longer. 
Sampling trajectories that involve slow molecular rearrangements and 
diffusive processes stands to benefit significantly from alternative 
methods of generating correlated trajectories.  

We consider three different ways to guide long trajectories:
introducing auxiliary forces; selecting among series of short trial
segments, as in Steered Transition Path Sampling
(STePS)~\cite{Guttenberg2012}; and advancing stochastic integrators
with correlated random numbers (which we refer to as ``noise
guidance'')~\cite{Crooks2001, Stoltz2007}.  
Of the three, only noise guidance yields an MC entropy production 
which is subextensive in the trajectory length.  
The other schemes, which accumulate extensive entropy production, 
cannot efficiently extend to sampling of long trajectories.  
Strong noise guidance is not, however, a panacea; correlated noises 
need not imply correlated trajectories.  We illustrate this point by 
considering Glauber dynamics of a two-dimensional Ising model and 
Langevin dynamics of a two-dimensional Weeks-Chandler-Andersen (WCA) 
fluid.  
Only when microscopic degrees of freedom have a small number of 
discrete possibilities, as in the lattice dynamics, is it possible 
to generate correlated long-timescale trajectories by tuning the noise.

The structure of the paper is as follows.  First we introduce and
discuss the perspective of MC moves as non-equilibrium processes which
produce entropy, detailing the consequences of constraints analogous to 
fluctuation theorems and the second law of thermodynamics.  
Next we review transition path sampling in stochastic dynamics and 
demonstrate the challenge posed by long trajectories in the context of
trajectory sampling of a one-dimensional random walker.  
We then analyze alternative strategies to correlate long trajectories 
of a one-dimensional single particle system and of a two-dimensional Ising
model.  
Finally, we explore how strongly noise guidance correlates trajectories 
in example systems, and then conclude.

\section{Markov Chain Monte Carlo Entropy Production}
\label{sec:MCEP}
We start by discussing a very general perspective on traditional
Metropolis-Hastings MCMC sampling~\cite{Metropolis1953,Hastings1970}.
Consider the problem of sampling a configuration, $\mathbf{x}$,
according to probability distribution $P(\mathbf{x})$.  
For example, $\mathbf{x}$ could be a vector of the coordinates and 
momenta of $N$ hard spheres, the state of spins in an Ising model, 
or all coordinates of a classical trajectory.  
The Metropolis-Hastings algorithm generates a Markov chain, which can be
thought of as a dynamics through configuration space with the
steady-state distribution $P(\mathbf{x})$.  
This dynamics obeys detailed balance but is not necessarily physical.

One typically splits each Monte Carlo move into two steps.
First, a change from $\mathbf{x}$ to a new state $\tilde{\mathbf{x}}$
is proposed according to a generation probability, 
$P_{\text{gen}}[\mathbf{x} \rightarrow \tilde{\mathbf{x}}]$.
Throughout this paper we will refer to $\mathbf{x}$ as a reference 
and $\tilde{\mathbf{x}}$ as the trial.
This trial is conditionally accepted with probability
\begin{equation}
P_{\text{accept}}[\mathbf{x} \rightarrow \tilde{\mathbf{x}}] = 
\min \left[1, e^{-\omega}\right],
\label{eq:Metropolis}
\end{equation}
where
\begin{equation}
\omega = \ln \frac{P(\mathbf{x}) P_{\text{gen}}[\mathbf{x} \rightarrow \tilde{\mathbf{x}}]}{P(\tilde{\mathbf{x}}) P_{\text{gen}}[\tilde{\mathbf{x}} \rightarrow \mathbf{x}]}.
\label{eq:entropyproduction}
\end{equation}
Together, these two steps ensure detailed balance, guaranteeing that
the equilibrium distribution $P(\mathbf{x})$ is stationary under the
MC protocol.  
Lacking the conditional acceptance step, such an MC procedure would 
generally drive a system away from its equilibrium distribution.  
We find it instructive to view this notional process as a genuine 
non-equilibrium transformation, one that would generate nonzero entropy
in most cases.  
In the formalism of stochastic thermodynamics, the resulting entropy 
production corresponds precisely to the quantity $\omega$ defined in
Eq.~\eqref{eq:entropyproduction}~\cite{Seifert2005}.  

The MC acceptance step effectively filters realizations of this 
non-equilibrium process, with a bias towards low values of $\omega$.
By construction, the bias exactly negates the tendency of trial move 
generation to drive a system out of equilibrium.
From this perspective, the countervailing tendencies of proposal and 
acceptance are akin to the operation of a Maxwellian demon, which by 
contrast filters realizations of \emph{equilibrium} dynamics with a 
bias that creates a \emph{non-equilibrium} state.

Viewing the procedure in the language of entropy production distributions 
reveals an important asymmetry of $P(\omega)$.
Following the more general demonstration of an entropy production 
fluctuation theorem~\cite{Crooks1999}, note that
\begin{align}
\nonumber P(\omega) &= \int d\mathbf{x} \ d\tilde{\mathbf{x}} \ P(\mathbf{x}) P_{\text{gen}}(\mathbf{x} \rightarrow \tilde{\mathbf{x}}) \delta(\omega - \omega(\mathbf{x}, \tilde{\mathbf{x}}))\\
\nonumber &= \int d\mathbf{x} \ d\tilde{\mathbf{x}} \ e^{\omega(\mathbf{x}, \tilde{\mathbf{x}})} P(\tilde{\mathbf{x}}) P_{\text{gen}}(\tilde{\mathbf{x}} \rightarrow \mathbf{x}) \delta(\omega + \omega(\tilde{\mathbf{x}}, \mathbf{x}))\\
&=e^\omega P(-\omega),
\label{eq:fluctthm}
\end{align}
with $\omega(\mathbf{x}, \tilde{\mathbf{x}})$ representing the entropy
produced by a proposal move from $\mathbf{x}$ to $\tilde{\mathbf{x}}$
and $\delta$ denoting the Dirac delta function.  
We are more likely to propose moves with positive entropy production 
than we are to choose their negative counterparts.  
The straightforward corollary, $\left<\omega\right> \geq 0$, is by 
analogy a statement of the second law, and the equality is satisfied 
if and only if $P(\omega) = \delta(\omega)$.  
A further consequence of Eq.~\ref{eq:fluctthm} relates the 
MC acceptance rate to the probability of attempting a move with
a negative value of $\omega$, which we call $p_{<}$.
Specifically,
\begin{equation}
\left<P_{\text{accept}}\right> = \int d\omega \ P(\omega) \min \left[1, e^{-\omega}\right] = 2 p_{<},
\end{equation}
which has been noted in the related context of replica exchange Monte
Carlo~\cite{Park2008}.
As $\left<\omega\right>$ increases, $p_{<}$, and
therefore $\left<P_{\text{accept}}\right>$, tends to decrease.
We will see that $\left<\omega\right>$ scales with the number of driven
degrees of freedom, such that Monte Carlo sampling of chain molecules 
or of long trajectories becomes especially challenging. 

We focus below on the sampling of dynamical pathways (rather than
individual configurations). 
In this case $\omega$ is an ``entropy production'' only by analogy, 
since the ``non-equilibrium transformations'' effected by MC trial moves
occur in the more abstract space of trajectories.
Lessons from Eq.~\eqref{eq:fluctthm} are nevertheless illuminating in
the context of this abstract space.

\section{Transition Path Sampling with Stochastic Dynamics}
\subsection{Trajectory Space and Trajectory Subensembles}

Let us now specialize to the sampling of discrete-time stochastic 
trajectories with a fixed number of steps, $t_{\text{obs}}$.
The probability of observing a trajectory, 
$\mathbf{x}(t) \equiv \left\{\mathbf{x}_0, \mathbf{x}_1, \hdots, \mathbf{x}_{t_{\text{obs}}}\right\}$, 
can be written as
\begin{equation}
P_{0}[\mathbf{x}(t)] \propto \rho_{\text{init}}(\mathbf{x}_0) \prod_{t = 0}^{t_{\text{obs}}-1} p(\mathbf{x}_t \rightarrow \mathbf{x}_{t+1}),
\label{eq:pathprob}
\end{equation}
where $\rho_{\text{init}}$ is a distribution for the initial time 
point, frequently an equilibrium or  steady state distribution.
The probability of each time propagation step is denoted 
$p(\mathbf{x}_t \rightarrow \mathbf{x}_{t+1})$, the form of which 
depends on details of the stochastic dynamics.
We refer to this propagation as the natural dynamics.
Representative trajectories can be generated by sampling the initial 
state and propagating natural dynamics.

In many contexts, it is useful to study a biased trajectory ensemble 
constructed to highlight particular rare events.
Common examples include the reactive subensemble,
\begin{equation}
P_{\text{reactive}}[\mathbf{x}(t)] \propto P_{0}[\mathbf{x}(t)] h_{\text{A}}(\mathbf{x}_0) h_{\text{B}}(\mathbf{x}_{t_{\text{obs}}}),
\label{eq:reactivepathprob}
\end{equation}
and the so-called tilted ensemble,
\begin{equation}
P_{\text{tilted}}[\mathbf{x}(t), s] \propto P_0[\mathbf{x}(t)] e^{-s K[\mathbf{x}(t)]}.
\label{eq:tiltedpathprob}
\end{equation}
In the former case, $h_{\text{A}}$ and $h_{\text{B}}$ are indicator 
functions which constrain the endpoints of the trajectory to fall in 
regions of phase space corresponding to reactants and products of a 
chemical reaction or other barrier crossing process~\cite{Bolhuis2002}.
In the latter case, $K[\mathbf{x}(t)]$ is an order parameter reporting 
on dynamical properties of the trajectory (e.g.\ the 
current~\cite{Derrida1998}, activity~\cite{Lecomte2007,Hedges2009}, or 
entropy production~\cite{Lebowitz1999,Mehl2008,Seifert2012}) and $s$ sets 
the strength of bias~\cite{Garrahan2009}.
These biased ensembles highlight classes of trajectories only rarely 
sampled by the natural dynamics.
To effectively sample them, a Markov chain of correlated trajectories 
is constructed.
The correlations between subsequent steps of the Markov chain ensure 
that newly-generated trajectories are likely to share the rare features 
that made the prior trajectory a good representative of the biased 
ensemble.

\subsection{Sampling with Shooting Moves}
\label{sec:shootingsec}
One of the most general and effective methods for generating a trial
trajectory is the shooting move, which is particularly well-suited to
sampling equilibrium dynamics~\cite{Bolhuis2002}.  
The move proceeds as follows.  
A discrete time, $t_{\text{shoot}}$, between $0$ and $t_{\text{obs}}$ 
is uniformly selected and designated the shooting time.  
The state of the system at $t_{\text{shoot}}$, perhaps slightly 
modified, is then propagated forward and backward in time with natural
dynamics to yield a trial trajectory, $\tilde{\mathbf{x}}(t)$.  
The probability of generating this trial takes the form
\begin{align}
\nonumber P_{\text{gen}}[\tilde{\mathbf{x}}(t)] \propto &\rho_{\text{init}}(\mathbf{x}_0) p_{\text{gen}}\left(\mathbf{x}_{t_{\text{shoot}}}\rightarrow \tilde{\mathbf{x}}_{t_{\text{shoot}}}\right) \times\\
&\prod_{t=0}^{t_{\text{shoot}}-1}\bar{p}(\mathbf{x}_{t+1}\rightarrow \mathbf{x}_{t}) \prod_{t=t_{\text{shoot}}}^{t_{\text{obs}}-1} p(\mathbf{x}_t \rightarrow \mathbf{x}_{t+1}),
\label{eq:shootpgen}
\end{align}
where $\bar{p}$ is the transition probability for time-reversed 
dynamics and $p_{\text{gen}}$ is the probability of the perturbation 
at the shooting time.
In the language of Section~\ref{sec:MCEP}, the entropy produced by 
this trial move is given by
\begin{align}
\nonumber \omega &= \ln \frac{\rho_{\text{init}}(\mathbf{x}_0) h_{\text{A}}(\mathbf{x}_0) h_{\text{B}}(\mathbf{x}_{t_{\text{obs}}}) p_{\text{gen}}(\mathbf{x}_{t_{\text{shoot}}} \rightarrow \tilde{\mathbf{x}}_{t_{\text{shoot}}})}{\rho_{\text{init}}(\tilde{\mathbf{x}}_0) h_{\text{A}}(\tilde{\mathbf{x}}_0) h_{\text{B}}(\tilde{\mathbf{x}}_{t_{\text{obs}}}) p_{\text{gen}}(\tilde{\mathbf{x}}_{t_{\text{shoot}}} \rightarrow \mathbf{x}_{t_{\text{shoot}}})}\\
& \ \ \ \ + \sum_{t = 0}^{t_{\text{shoot}} - 1} \ln \frac{p(\mathbf{x}_t\rightarrow \mathbf{x}_{t+1}) \bar{p}(\tilde{\mathbf{x}}_{t+1} \rightarrow \tilde{\mathbf{x}}_t)}{\bar{p}(\mathbf{x}_{t+1}\rightarrow \mathbf{x}_{t}) p(\tilde{\mathbf{x}}_{t} \rightarrow \tilde{\mathbf{x}}_{t+1})}.
\label{eq:shootingep}
\end{align}
For long trial trajectories to be accepted by the MCMC scheme, $\omega$ 
must be small.
However, when $p$ and $\bar{p}$ are not equal (as is the case in driven 
processes), the sum in Eq.~\eqref{eq:shootingep} has order 
$t_{\text{obs}}$ nonvanishing terms~\footnote{Since $t_{\text{shoot}}$ 
is uniformly selected between $0$ and $t_{\text{obs}}$, 
$t_{\text{shoot}}$ is of order $t_{\text{obs}}$.  
Furthermore, while neighboring terms in the sum may be correlated, 
there will still be order $t_{\text{obs}}$ independent terms for 
sufficiently long $t_{\text{obs}}$.}.
Consequently, $\left<\omega\right>$ scales linearly with 
$t_{\text{obs}}$, and $P(\omega)$ adopts the long-time form
\begin{equation}
P(\omega) \sim \exp\left[-t_{\text{obs}} I(\omega / t_{\text{obs}})\right],
\label{eq:ldf}
\end{equation}
with large deviation rate function $I(\omega / t_{\text{obs}})$.  
From this asymptotic expression for the entropy produced by a TPS move, 
one might generally expect that the corresponding acceptance rate
decreases exponentially as $t_{\text{obs}}$ grows long.

This extensive growth of $\left<\omega\right>$ with time has an 
important and general exception, namely the case of microscopically 
reversible dynamics.
Under those conditions, the sum in Eq.~\eqref{eq:shootingep} vanishes 
and the only entropy production is contributed from the endpoints of 
the trajectory (e.g., $h_{\text{A}}$ and $h_{\text{B}}$). 
Since this entropy production is subextensive in time, long 
trajectories appear no more difficult to sample than short ones.  
The acceptability of trial trajectories, however, is also subject to 
biases like those expressed in Eqs.~\eqref{eq:reactivepathprob} 
and~\eqref{eq:tiltedpathprob}.  
Because long trajectories typically decorrelate strongly from one 
another, the rare, biased qualities of a reference trajectory 
(e.g., reactivity or inactivity) are recapitulated in the trial 
path with a probability that also decays with $t_{\text{obs}}$.

We conclude that the challenges for efficiently sampling long
trajectories are twofold.  
The TPS move must produce entropy that is subextensive in observation 
time or the method will not scale to long trajectories.  
Additionally, one must preserve strong correlations between 
$\mathbf{x}(t)$ and $\tilde{\mathbf{x}}(t)$, so that 
rare properties of interest are retained in the trial trajectory.
In the next section we show that these two goals are often conflicting.  
In particular, we examine three general schemes for engineering 
correlations between reference and trial trajectories in 
shooting-like moves.  
Two of the schemes fail to exhibit subextensive entropy production 
scaling while the remaining scheme can only maintain strong trajectory 
correlations in special cases.

\section{Guided Dynamics of a 1d Random Walker}
\begin{figure*}[ht]
\centering
\includegraphics[width=0.9\textwidth]{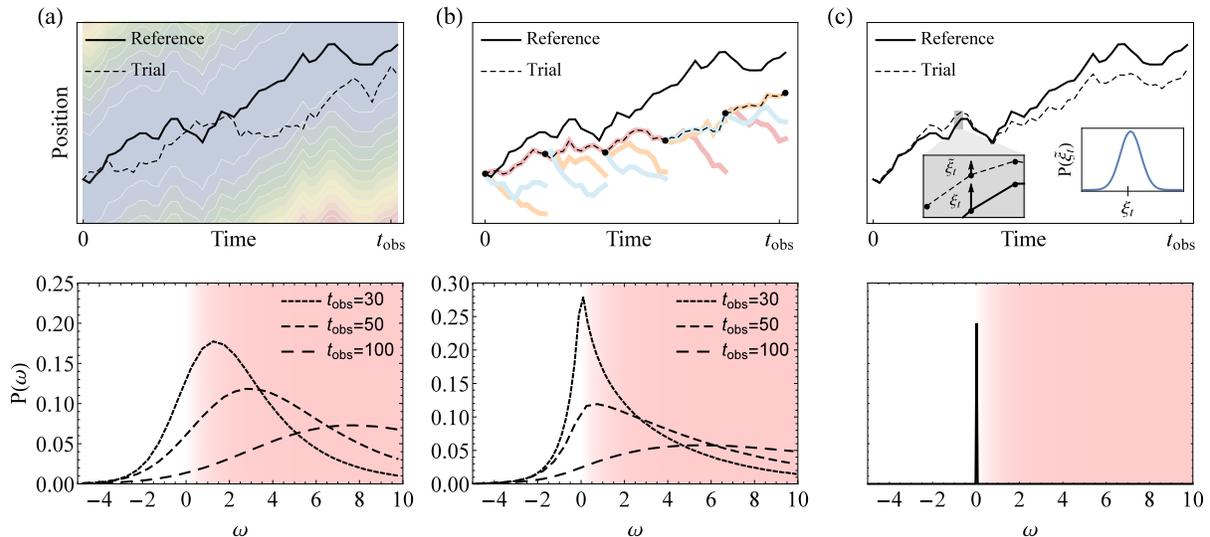}
\caption[Three Strategies for Guiding Trajectories] 
{
  Three guidance schemes for generating a trial trajectory that
  maintains proximity to a reference trajectory. For the specific case
  of a one-dimensional random walker, upper panels illustrate the
  consequences of (a) artificial corralling forces (b) preferential 
  selection of short trial branches and (c) correlated noise histories.  
  Bottom panels show the corresponding distributions of trajectory space 
  entropy production $\omega$.
  The intensity of red shading reflects the probability that trial moves
  are rejected.  
  For cases (a) and (b), the average entropy production is nonzero and 
  grows with trajectory length $t_{\text{obs}}$.
  With an appropriately designed noise guidance scheme (c), symmetric
  selection of noise variables results in identically zero entropy
  production for all trajectory lengths.
}
\label{fig:guidancefig}
\end{figure*}

We explore the three methods for trial trajectory generation in the 
specific context of a one-dimensional discrete-time random walker with 
equation of motion
\begin{equation}
x_{t+1} = x_t + \xi_{t},
\label{eq:1deom}
\end{equation}
where at the $t^{\text{th}}$ timestep, the noise $\xi_t$ is drawn 
from the normal distribution with zero mean and variance $\sigma^2$.  
As a simple illustration focusing on the effects of entropy production,
suppose we want to sample the unbiased trajectory distribution
\begin{equation}
P_0[x(t)] \propto 
\delta(x_0)
\exp\left[-\sum_{t=0}^{t_{\text{obs}}-1} \frac{(x_{t+1}-x_t)^2}{2\sigma^2}\right],
\end{equation}
where the initial position is set to zero without loss of 
generality~\footnote{Typically TPS procedures concurrently sample 
the initial configuration, but that complication is not necessary 
for our illustration.}.
To construct a reference trajectory $x(t)$, we draw a value for 
$\xi_t$ at each timestep and propagate the walker's position 
according to Eq.~\eqref{eq:1deom}.  
A trial trajectory is then generated by evolving dynamics from the 
same initial configuration (with a different realization of the noise 
or perhaps even a different equation of motion).

We imagine that it is desirable for the trial trajectory to retain a
significant correlation with the reference path. 
This goal is motivated by the challenges of sampling biased ensembles 
as discussed above, but for the sake of simplicity we do not include 
such a bias here. 
To ensure this correlation, we employ shooting moves that differ from 
the conventional procedure described in Section~\ref{sec:shootingsec}. 
Specifically, we implement and scrutinize three distinct ways to
engineer correlation over long times: (a) adding artificial forces 
that pull the trial trajectory closer to the reference, (b) 
preferentially selecting among sets of otherwise unbiased short path 
segments, or (c) using correlated histories of noises.  
We assess the influence of these three biasing methods on
the MCMC efficiency by characterizing the distribution $P(\omega)$.

\subsection{Guiding Forces}
\label{sec:guidingforces}
We first consider effecting correlations with guiding forces, 
i.e., artificial contributions to the effective potential that tend to 
lead the trial trajectory toward the reference.  
This strategy is equivalent to using steered molecular 
dynamics~\cite{Isralewitz2001} to generate new trajectories.  
The trial trajectory $\tilde{x}(t)$ is grown with the equation of motion
\begin{equation}
\tilde{x}_{t+1} = \tilde{x}_t + \tilde{\xi}_{t} + k(x_t - \tilde{x}_t).
\label{forceeom}
\end{equation}
We denote $\tilde{\xi}_t$ as the trial trajectory noise at timestep 
$t$, also drawn from a Gaussian with mean zero and variance $\sigma^2$.
The linear spring constant $k$ adjusts the strength of correlation 
between reference and trial trajectories.
The probability that this guided dynamics generates a particular 
trial from the reference is given by
\begin{equation}
P_{\text{gen}}[x(t) \rightarrow \tilde{x}(t)] \propto \exp\left[-\sum_{t=0}^{t_{\text{obs}}-1} \frac{\left(\tilde{x}_{t+1} - \tilde{x}_t - k (x_t - \tilde{x}_t)\right)^2}{2 \sigma^2}\right].
\label{eq:forcePgen}
\end{equation}
The entropy production associated with the trial move depends also on 
the probability of generating the reverse TPS move, growing the 
reference trajectory with extra forces pulling it close to the trial.
It is straightforward to compute $\omega$ from 
Eq.~\eqref{eq:entropyproduction},
\begin{equation}
\omega = -\frac{k}{\sigma^2} \sum_{t=0}^{t_{\text{obs}}-1} (x_{t} - \tilde{x}_{t})(x_{t+1} +\tilde{x}_{t+1} - x_{t} - \tilde{x}_{t}).
\label{eq:positionentropyproduction}
\end{equation}
In this approach, $\omega$ can be physically interpreted as the 
difference between two work values: that expended by the artificial 
force to guide the trial trajectory, versus the work that would be 
required to conversely guide the reference.
The resulting distribution of entropy production, obtained
from numerical sampling, is shown in in Fig.~\ref{fig:guidancefig}(a).

Since $\omega$ is given by a sum over all $t_{\text{obs}}$ timesteps, 
$P(\omega)$ adopts a large deviation form as in Eq.~\eqref{eq:ldf}, 
and $\left<\omega\right> \propto t_{\text{obs}}$.
These scaling properties are demonstrated numerically in 
Fig.~\ref{fig:ldffig}(a) and analytically in 
Appendix~\ref{app:guidingforces}.
\begin{figure*}[ht]
\centering
\includegraphics[width=0.95\textwidth]{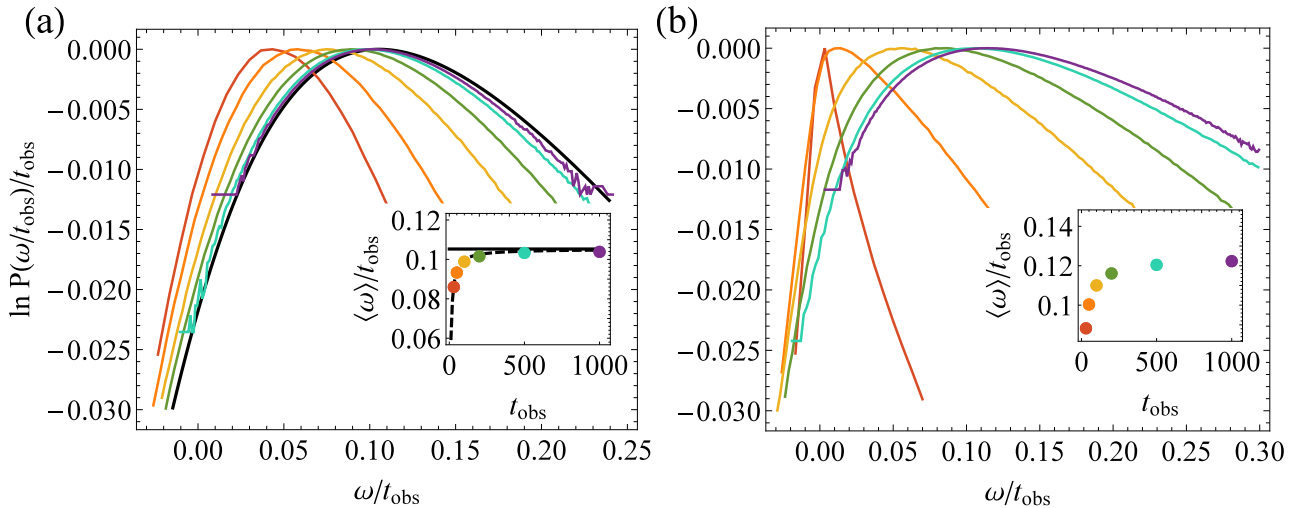}
\caption[Entropy Production Large Deviation Functions for Guided
  Trajectories]
{
  Entropy production statistics for trial moves with guiding forces 
  (a) and guiding choices (b), as discussed in 
  Sections~\ref{sec:guidingforces} and~\ref{sec:guidingchoices}
  of the main text.  
  Results in (a) are shown for $k = 0.1$; black curves indicate 
  the long-time behavior determined analytically in 
  Appendix~\ref{app:guidingforces}.  
  Results in (b) are shown for $n = 3, \tau = 10$ and with 
  $f(x) = e^{-|x|}$.
  In both panels, different colors indicate different trajectory lengths,
  ascending from left to right: $t_{\text{obs}} = 30$ (red), 
  50 (orange), 100 (yellow), 200 (green), 500 (cyan), and 1000 (purple).
}
\label{fig:ldffig}
\end{figure*}
In the appendix we re-express $\omega$ in terms of the $\xi$ and 
$\tilde{\xi}$ variables, which can be integrated over to yield
\begin{equation}
\left<\omega\right> = \frac{2}{(k-2)^2}\left[\left(2-k\right)k t_{\text{obs}}-1 + (k-1)^{2t_{\text{obs}}}\right].
\label{eq:meanscaling}
\end{equation}
Indeed, for $0 < k < 2$, this expression gives the anticipated long 
time scaling with $t_{\text{obs}}$,
\begin{equation}
\left<\omega\right> \sim \frac{2kt_{\text{obs}}}{2-k}.
\label{eq:meanscalinglongtime}
\end{equation}
As seen in Fig.~\ref{fig:guidancefig}(a), the negative-$\omega$ tail 
of $P(\omega)$, which gives rise to MCMC acceptances, becomes 
correspondingly small for large $t_{\text{obs}}$.

\subsection{Guiding Choices}
\label{sec:guidingchoices}
In both Sections~\ref{sec:shootingsec} and~\ref{sec:guidingforces} we
showed that time-extensive entropy production arises generically when
we do not use natural (forward) dynamics to generate a trajectory.
Dynamical biases can alternatively be achieved by preferentially
selecting among different examples of natural dynamics. 
At a high level, conventional TPS~\cite{Bolhuis2002} is just such 
an approach, constructing biased trajectory ensembles through selection 
rather than artificial forces. 
Can this strategy be used effectively to impose resemblance between 
reference and trial trajectories?

We consider a scheme very similar in spirit to the STePS
algorithm~\cite{Guttenberg2012}.
Like configurational-bias MC sampling of a polymer~\cite{Frenkel2001}, 
the STePS procedure generates a long trajectory by piecing together 
short segments, as illustrated in Fig.~\ref{fig:guidancefig}(b).  
To generate a new segment, one starts at the end of the previous 
segment and samples a collection of short, unbiased trajectories 
according to the natural dynamics, which we will refer to as branches.  
One of these branches is selected as the next segment of the trial 
trajectory, with a preference for branches that stay close to the 
reference trajectory. 
(Proximity could be judged in different ways, e.g., through Euclidean
distance in the full phase space, or with respect to an order
parameter).  
Though each branch is grown with natural dynamics, the added segment 
is biased.  
To show that this bias affects acceptance rates in the same manner 
as the guiding forces bias, we compute the entropy produced by a TPS move.

Starting at the initial condition of the reference trajectory, we 
grow $n$ branches of length $\tau$ according to
\begin{equation}
x_{t+1}^{(\alpha)} = x_t^{(\alpha)} + \xi_t^{(\alpha)},
\label{eq:branches}
\end{equation}
where $\alpha$ is an index over the $n$ independent samples of the 
natural dynamics.
Of these $n$ possibilities for the $i^{\text{th}}$ segment of the trial 
trajectory, we select branch $\alpha$ with probability
\begin{equation}
P_{\text{select}}(\alpha) = \frac{f(|x_{i\tau}^{(0)} - x_{i\tau}^{(\alpha)}|)}{\sum_{\gamma=1}^n f(|x_{i\tau}^{(0)} - x_{i\tau}^{(\gamma)}|)},
\end{equation}
where $f$ is a weighting function with a maximum when its argument is 
zero, for example a Gaussian centered on zero.
The reference trajectory is indicated by a superscript $(0)$.
Starting from the end of the chosen branch, the growth procedure is 
repeated with $n$ new branches of length $\tau$.

While each time propagation step uses segments of unbiased natural 
dynamics, the selection of preferred branches exerts a bias which 
ultimately leads to a nonvanishing entropy production,
\begin{equation}
\omega = - \sum_{i=1}^{t_{\text{obs}}/\tau} \ln \frac{\sum_{\gamma \neq 0} f(|x_{i \tau}^{(\gamma)} - x_{i \tau}^{(0)}|)}{\sum_{\gamma \neq \alpha_i} f(|x_{i \tau}^{(\gamma)} - x_{i \tau}^{(\alpha_i)}|)},
\end{equation}
where $\alpha_i$ is the index of the selected branch for the 
$i^{\text{th}}$ segment.
The calculation of this entropy production requires generation of 
the backwards TPS move, in which the $(0)$ branch is always selected.

In the preceding section we discussed that the entropy produced by 
guiding forces could be thought of in terms of a work performed by 
the bias.
From that perspective, this guiding choices scheme trades work for 
information.
We bias the dynamics not by applying explicit forces but instead by 
selecting the preferred branches based on information about the 
likelihood that a branch stays close to the reference.
In particular, $\omega$ is a difference between the Shannon information 
associated with selecting the set of trial branches which produced the 
trajectory $\tilde{x}$(t) and the information associated with selecting 
the reference branches in a reverse TPS move.
As with biasing forces, the trajectory space entropy production exhibits 
large deviation scaling with $\left<\omega\right> \propto t_{\text{obs}}$.
Numerical demonstrations of this scaling are provided in 
Fig.~\ref{fig:ldffig}(b).
Consequently, acceptance probabilities drop precipitously in the long 
time limit.

\subsection{Guiding Noises}

As a third scheme for engineering path correlation, we consider 
generating a trial trajectory with natural dynamics but with biased 
noises.  
Rather than trying to corral trajectories to proceed along similar 
paths, one may impose much simpler correlations between their 
underlying noise histories~\cite{Crooks2001}.  
Consider a TPS trial move which consists of re-propagating dynamics 
from the initial timestep using new noises $\tilde{\xi}$ that differ 
only slightly from the old noises $\xi$,
\begin{equation}
\tilde{\xi}_t = \alpha \xi_t + \sqrt{1-\alpha^2} \eta_t,
\label{eq:stoltzalpha}
\end{equation}
where $\eta_i$ is sampled from a Gaussian distribution with zero mean
and variance $\sigma^2$.
In Section~\eqref{sec:guidingchoices}, the symbol $\alpha$ was used as an
index, but here we redefined $\alpha$ to be the parameter controlling
noise correlations.
Guiding Gaussian noise variables in this manner has been referred to as 
a Brownian tube proposal move~\cite{Stoltz2007}.  Unlike the prior two
kinds of moves, the Brownian tube proposal produces strictly vanishing
entropy production $\omega$ for all trials regardless of trajectory
length.
The cancellation results from some algebra after writing the path weights 
and generation probabilities in terms of the noise variables,
\begin{equation}
\frac{P({\boldsymbol \xi}) P_{\text{gen}}({\boldsymbol \xi} \rightarrow \tilde{{\boldsymbol \xi}})}{P(\tilde{{\boldsymbol \xi}})P_{\text{gen}}(\tilde{{\boldsymbol \xi}} \rightarrow {\boldsymbol \xi})} = \frac{\exp\left[{-\frac{{\boldsymbol \xi}^2}{2\sigma^2}}\right] \exp\left[{-\frac{(\tilde{{\boldsymbol \xi}}-\alpha {\boldsymbol \xi})^2}{2(1-\alpha^2)\sigma^2}}\right]}{\exp\left[{-\frac{\tilde{{\boldsymbol \xi}}^2}{2\sigma^2}}\right] \exp\left[{-\frac{({\boldsymbol \xi}-\alpha \tilde{{\boldsymbol \xi}})^2}{2(1-\alpha^2)\sigma^2}}\right]}=1,
\label{eq:noise}
\end{equation}
where $\boldsymbol \xi$ is a vector detailing noises at all times.

The vanishing entropy production is achieved by \emph{independently
sampling} the noise variables.
In the previously discussed approaches, the bias applied to any one noise 
variable depended on how far astray the trial trajectory had drifted from 
its reference up to that point in time.
Such history-dependent biasing coupled the sampling of one noise variable 
to all of the previous noises, ultimately giving rise to the nonvanishing 
$\omega$.
By sampling all noises independently, we can perturb the $\xi$ variables 
in a symmetric manner.
For noises drawn from a Gaussian distribution, this symmetric perturbation 
was given in Eq.~\eqref{eq:stoltzalpha}, but the strategy of symmetrically 
sampling independent noises generalizes to other choices of stochastic 
dynamics.
For example, Hartmann has applied these methods with uniform random 
variable noises to Monte Carlo dynamics in the form of Wolff dynamics 
of a two-dimensional Ising model~\cite{Hartmann2014}.

\begin{figure*}[ht]
\centering
\includegraphics[width=0.85\textwidth]{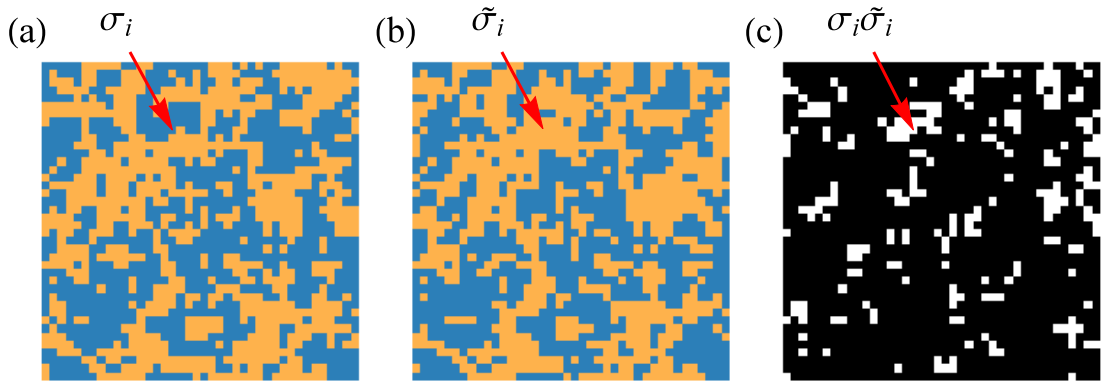}
\caption[Overlap Between Different Ising Trajectories]
{
  Correlations between a reference trajectory and a trial trajectory 
  generated by the noise guidance method described in 
  Section~\ref{sec:isingdyn} using 
  $\epsilon_{\text{dir}} = \epsilon_{\text{site}} = 10^{-3}$ and 
  $\epsilon_{\text{acc}} = 0.1$. 
  The two trajectories begin with identical initial conditions and evolve
  through 100 sweeps of push up/push down Monte Carlo steps at 
  $\beta J = 0.3, h = 0$.  
  Final configurations of reference and trial trajectories are shown 
  in (a) and (b), respectively. 
  The site-wise overlap between these two configurations is depicted 
  in (c), where black indicates spin alignment and white indicates 
  anti-alignment.
}
\label{fig:latticefig}
\end{figure*}

Using correlated noise histories to sample Monte Carlo trajectories 
avoids the time-extensive bias that arose from guiding paths in 
configuration space, but this merit comes at a cost.
If the reactive or tilted ensembles are to be sampled, it is important 
that the guidance scheme produces highly correlated trajectories.
That is to say, the $x$ coordinates, not just the $\xi$ coordinates, 
must be correlated.
When will similar noise histories produce similar trajectories?
In the remainder of the paper we address this question in the context 
of two dynamical systems, one on-lattice and the other off-lattice.

\section{Efficacy of Noise Guidance}

In the preceding section we noted that sampling trajectories with 
noise-guided shooting moves avoids a time-extensive MC entropy production.
However, we seek correlated trajectories, not just correlated noises.
When trajectories with correlated noises synchronize, efficient path 
sampling of long trajectories can be achieved.
But under what conditions should such synchronization be expected?
We investigate this question by studying lattice dynamics of a 
two-dimensional Ising model and off-lattice dynamics of a WCA fluid, 
also in two dimensions.
We show that synchronization can be achieved with a suitable treatment 
of Ising dynamics.
This success does not extend to our example of off-lattice dynamics.

\subsection{Ising Dynamics}
\label{sec:isingdyn}

Let us first consider a two-dimensional Ising model consisting of $N$ spins.
The $i^{\text{th}}$ spin, denoted $\sigma_i$, takes the value $\pm 1$.
The lattice evolves, at inverse temperature $\beta$,  under single 
spin-flip Glauber dynamics with Hamiltonian
\begin{equation}
H = -h \sum_i \sigma_i - J \sum_{\left<ij\right>} \sigma_i \sigma_j.
\label{eq:ising}
\end{equation}
The spins interact in the usual Ising manner; they couple to nearest 
neighbors with coupling constant $J$ and to an external field $h$.
Each spin-flip trial move requires us to choose two random numbers 
uniformly from $[0, 1)$.
One random number, $\xi_{\text{site}}$, determines which site will be 
flipped.
The other random number, $\xi_{\text{acc}}$, determines whether to accept 
or reject the flip.
Given $\xi_{\text{site}}$ and $\xi_{\text{acc}}$, the spin-flip move is 
deterministic:
\begin{enumerate}
\item Choose spin $i$ = ceiling$(\xi_{\text{site}} N)$ to act on.
\item Construct a trial state by flipping spin $i$.
\item Compute the energy difference, $\Delta E$, between the original configuration and the trial.
\item Accept the spin flip if $\xi_{\text{acc}} < (1 + \exp(\beta \Delta E))^{-1}$.
\end{enumerate}
By carrying out $t_{\text{obs}}$ sequential MC sweeps, each consisting of 
$N$ spin flip moves, we construct an Ising trajectory, $\sigma(t)$.
The effective unit of time is thus taken to be a MC sweep.

Now consider a noise-guided trial TPS move designed to generate a 
trajectory $\tilde{\sigma}(t)$ which is correlated with $\sigma(t)$.
At every MC step we alter $\xi_{\text{site}}$ and $\xi_{\text{acc}}$ 
to some trial values, $\tilde{\xi}_{\text{site}}$ and 
$\tilde{\xi}_{\text{acc}}$.
There is significant freedom in doing so, while producing zero 
entropy~\footnote{We want symmetric proposal probabilities, 
  $p(\xi \to \xi') = p(\xi' \to \xi)$.}, and we analyze one particular 
choice.
We focus first on updating the noise that chooses which spin 
to flip.
With probability $1-\epsilon_{\text{site}}$ we reuse the old noise, 
i.e., $\tilde{\xi}_{\text{site}} = \xi_{\text{site}}$.
Otherwise, we uniformly draw a new value of $\tilde{\xi}_{\text{site}}$ 
from the unit interval.
The tunable parameter $\epsilon_{\text{site}}$ controls the correlation 
between noise histories of the reference and trial trajectory.
We update the noises that control conditional acceptance, 
$\xi_{\text{acc}}$, in an analogous manner.
Another parameter, $\epsilon_{\text{acc}}$, is the probability of drawing 
new noise for $\tilde{\xi}_{\text{acc}}$.

\begin{figure*}[ht]
\centering
\includegraphics[width=0.9\textwidth]{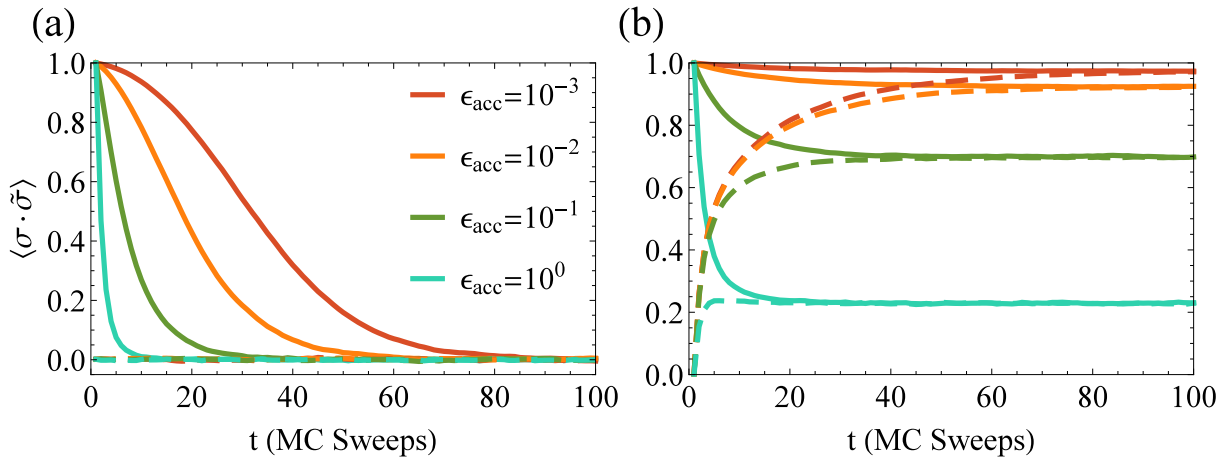}
\caption[Time-dependent Correlations in Guided Ising
  Trajectories]
{
  Average overlap between reference and trial trajectories of a 
  $40\times 40$ two-dimensional Ising model with $\beta J = 0.3$.  
  Results are shown in (a) for ordinary Glauber spin flip dynamics, 
  and in (b) for the modified directional dynamics described in 
  Section~\ref{sec:isingdyn}.  
  Two trajectories with identical initial conditions and 
  different-but-correlated noise histories (solid lines) maintain 
  a nonzero steady state overlap at long times only for the case of 
  push up/push down dynamics. 
  The same steady state values are obtained when the two trajectories 
  evolve from very different initial conditions (dashed lines) generated 
  by independently assigning each spin at random.  
  Different colors indicate different values of the noise guidance 
  parameter $\epsilon_{\text{acc}}$. 
  Ensemble-averaged results are shown for 
  $\epsilon_{\text{site}} = \epsilon_{\text{dir}} = 0.001$, with
  averages performed over 500 independent pairs of trajectories.
}
\label{fig:decorrelationfig}
\end{figure*}

Starting with the initial configuration of $\sigma(t)$, we construct 
$\tilde{\sigma}(t)$ by performing spin flips with the new trial noise 
history.
The trial and reference trajectories start in identical configurations, 
but we expect the correlation to decay as MC time advances.
To monitor the similarity between reference and trial, we
study the site-wise product between $\sigma$ and $\tilde{\sigma}$ as 
illustrated in Fig.~\ref{fig:latticefig}.
The average of this product over all spins,
\begin{equation}
\sigma \cdot \tilde{\sigma} = \frac{1}{N} \sum_{i=1}^N \sigma_i \tilde{\sigma}_i
\label{eq:sitewise}
\end{equation}
is a measure of correlation between $\sigma$ and $\tilde{\sigma}$.
Decorrelated configurations return a value of zero while identical
configurations return one.  Fig.~\ref{fig:decorrelationfig}(a) shows
that the correlation between $\sigma(t)$ and $\tilde{\sigma}(t)$
decays to zero at long times.  The rate of this decay is tuned by
$\epsilon_{\text{acc}}$ and $\epsilon_{\text{site}}$, the parameters
controlling the extent of noise correlation.  
At long times $\left<\sigma \cdot \tilde{\sigma}\right>$ eventually 
approaches zero, even with the strongest noise guidance.  
The corresponding ``uncorrelated'' configurations, however, bear a 
subtler resemblance.
Some regions are significantly correlated, while others are
significantly anti-correlated, averaging to give 
$\sigma \cdot \tilde{\sigma} \approx 0$.

Motivated by this subtler resemblance, we introduce a minor alteration
in the implementation of Glauber spin-flip dynamics. 
Below we detail this modification and show that it does in fact enable 
the preservation of correlation between trajectories over very long times.
As observed in the context of damage spreading, different choices of
Ising model Monte Carlo dynamics can result in identical equilibrium
states yet different dynamical properties~\cite{Hinrichsen1997,Mariz1990}.  
The effectiveness of correlated noises in guiding trajectories is, in 
effect, one such dynamical property.

In particular, we replace step 2 of the spin-flip move to include a
directionality.  We introduce another random number, 
$\xi_{\text{dir}} \in [0, 1)$, used to decide the trial state.
If $\xi_{\text{dir}} <0.5$, the trial state is a down spin; 
otherwise it is up.  
Whereas the conventional trial move effects an attempted spin flip, 
this trial move can be viewed as an attempt to push the spin either 
up or down, depending on the state of $\xi_{\text{dir}}$.  
As with the other noise variables, correlations between 
$\xi_{\text{dir}}$ and $\tilde{\xi}_{\text{dir}}$ are tuned by the 
probability $\epsilon_{\text{dir}}$ of resampling the noise.

The addition of directionality to the spin-flip dynamics results in 
moves which are trivially ineffectual.
For example, half of the spin moves attempt to ``push up'' a spin 
which is already up.
These moot moves are absent from the traditional implementation of 
single-spin-flip Glauber dynamics, which attempts a spin flip at every 
step of MC time.
In every other respect, the two schemes generate Markov chains with 
identical statistics.
They can therefore be made identical by excising moot moves, or, 
equivalently on average, by scaling time by a factor of two.

As Fig.~\ref{fig:decorrelationfig}(b) illustrates, the push up/push
down implementation of single-spin-flip Glauber dynamics allows the
trial trajectories, $\tilde{\sigma}(t)$ to remain tunably close to
$\sigma(t)$ for long times.  
By incorporating information about spin change directionality into 
the noise history, the noises signal not just how likely a spin is to
change, but in what direction it will change.  
Appropriately chosen $\epsilon$ parameters can create trajectories 
which remain tunably close to each other for arbitrarily long times.  
When averaged over the whole lattice, steady state correlation is 
maintained, but the correlations are not spatially homogeneous.  
As MC time progresses, the regions in which two trajectories are 
highly correlated move throughout the lattice, ensuring ergodic 
exploration of the trajectory space.

\begin{figure*}[ht]
\centering
\includegraphics[width=0.97\textwidth]{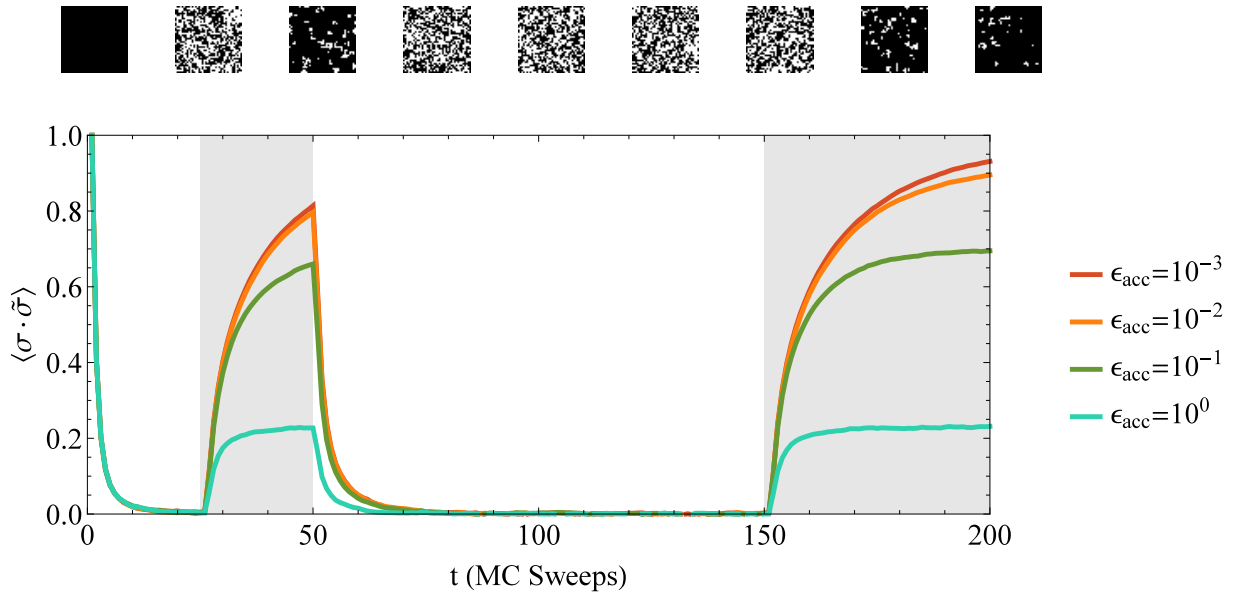}
\caption[Strategy for Guiding Lattice Dynamics with Important
  Intermediate States]
{ 
  Correlation between reference and trial trajectories of a 
  $40\times 40$ Ising model with $\beta J = 0.3$.
  Plotted lines are averages over $500$ independent pairs of 
  trajectories that evolve by push up/push down dynamics.
  Noise histories of each pair are generated in a correlated manner with
  $\epsilon_{\text{site}} = \epsilon_{\text{dir}} = 10^{-3}$
  during the intervals $t=25-50$ and $t=150-200$ (the shaded regions);
  noise guidance is absent at all other times.  
  The site-wise correlation between an example trajectory pair
  (with $\epsilon_{\text{acc}}=10^{-3}$) is shown above the graph, 
  with time advancing from left to right and adjacent configurations 
  separated by 25 MC sweeps.
}
\label{fig:intermediates}
\end{figure*}

For push up/push down dynamics, noise guidance does not merely
preserve correlations that existed at time zero.  We find that
correlated noise histories can in fact {\em induce} synchronization
between trajectories.  
To illustrate this synchronization effect, we have characterized the 
correlation between trajectories that share similar noise histories 
only intermittently. 
As shown in Fig.~\ref{fig:intermediates}, such paths acquire similarity 
during periods of strong noise guidance.  
This similarity degrades during periods without noise guidance, but can 
be recovered by re-introducing guidance, regardless of how significantly 
correlations have decayed.
Indeed, even very different initial configurations, propagated with
correlated noises, become more similar with time, their 
ensemble-averaged correlation
$\left<\sigma \cdot \tilde{\sigma}\right>$ converging to the same value as
for trajectories that are identical at time zero. 

\begin{figure*}[hbt]
\centering
\includegraphics[width=0.97\textwidth]{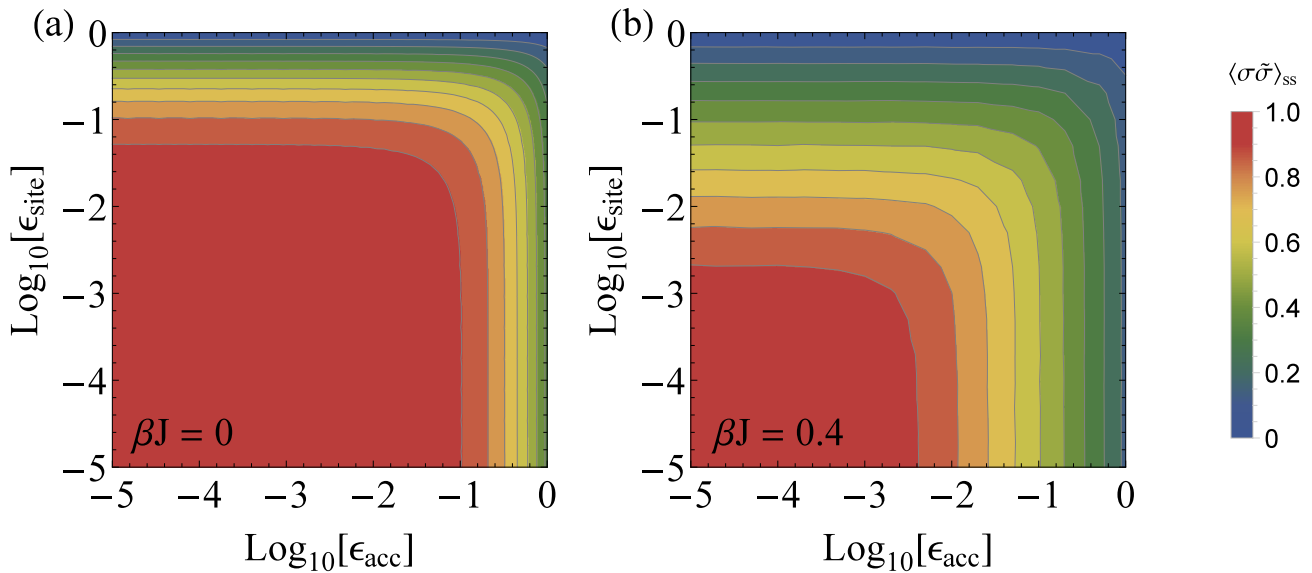}
\caption[Steady State Correlations in Guided Ising Trajectories]
{
  Steady state correlations between reference and trial trajectories of
  a $40\times 40$ Ising model, as a function of noise guidance strength. 
  Results are shown for push up/push down dynamics with
  $\epsilon_{\text{dir}} = 0.001$.  
  The high-temperature limit Eq.~\ref{eq:hightemp} is shown in (a). 
  Finite temperature behavior (b) was obtained by sampling 
  $4 \times 10^4$ pairs of trajectories for 340 MC sweeps each, 
  with $\beta J = 0.4$.
}
\label{fig:ssfig}
\end{figure*}

A nonzero steady state value of $\left<\sigma \cdot \tilde{\sigma}\right>$
is the quantitative signature of synchronization.  
The origins of this finite asymptotic correlation are transparent in 
the limit of weak coupling, $\beta J=0$. 
With the additional simplification $h=0$, each attempted spin flip is 
accepted with probability $1/2$ based on the value of $\xi_{\text{acc}}$, 
regardless of the states of neighboring spins.  
In this case the steady state overlap can be calculated analytically. 
To do so, we derive an equation of motion for the probability 
$p(\tau)$ that a given spin has identical values in the reference and 
trial trajectories after $\tau$ MC steps.
Note that $\tau$ differs from time $t$ by a factor of $N$.
The long-time, steady-state limit of this time evolution, 
$p_{\text{ss}} = \lim_{\tau \rightarrow \infty} p(\tau)$
, yields 
$\left<\sigma \cdot \tilde{\sigma}\right>_{\text{ss}} = 2 p_{\text{ss}}-1$.  
In Appendix~\ref{app:overlap}, we tabulate the various ways that a
selected spin can become identical in reference and trial trajectories
after a single timestep. 
From this enumeration, and the corresponding probabilities, we find
\begin{align}
\nonumber p(\tau+1) &= \frac{N-1}{N}p(\tau) - \frac{\epsilon_{\text{dir}}}{4N} \left(1-\epsilon_{\text{site}} + \frac{\epsilon_{\text{site}}}{N}\right)\left(1-\frac{\epsilon_{\text{acc}}}{2}\right)\\
& \  + \frac{1}{2N} \left(1-\frac{\epsilon_{\text{acc}}}{2}\right)\left(1-\epsilon_{\text{site}}+\frac{\epsilon_{\text{site}}}{N}\right) p(\tau) + \frac{1}{2N}.
\label{eq:eom}
\end{align}

The various terms of Eq.~\ref{eq:eom} describe the different ways that
a single MC move can impact the state of an arbitrarily chosen spin in
trial and reference trajectories.  
Since each move of our MC dynamics acts on a single site of the lattice, 
some moves do not involve the tagged spin at all, but instead some 
other lattice site; the first term in Eq.~\ref{eq:eom} reflects this 
possibility.
The second term accounts for the decrease in overlap when reference and 
trial trajectories accept a spin-flip at the same site but in opposite 
directions.  
The third term results from the constructive action of correlated noises
on the tagged spin, either maintaining existing correlation or inducing 
synchronization, as detailed in Appendix~\ref{app:overlap}.  
The final term accounts for random alignment of the tagged spin despite
uncorrelated noise variables, a possibility particular to degrees of 
freedom with a limited number of discrete states.
Equating $p(\tau)$ and $p(\tau+1)$ gives the steady state probability,
\begin{equation}
p_{\text{ss}} = \frac{1 - \frac{\epsilon_{\text{dir}}}{2} \left(1 - \frac{\epsilon_{\text{acc}}}{2}\right)\left(1 - \epsilon_{\text{site}} + \frac{\epsilon_{\text{site}}}{N}\right)}{2 - \left(1 - \frac{\epsilon_{\text{acc}}}{2}\right)\left(1 - \epsilon_{\text{site}} + \frac{\epsilon_{\text{site}}}{N}\right)}.
\label{eq:hightemp}
\end{equation}

Analytically calculating steady-state overlap at finite temperature
is not straightforward.
Numerical results, shown for $\beta J = 0.4$ in Fig.~\ref{fig:ssfig},
indicate that the dependence of overlap on strengths of noise
perturbation is generically similar to the $\beta J=0$ case analyzed
above.
Increasing $\beta J$ from zero does, however, slow the rate of
convergence to the steady state, while decreasing the degree of steady 
state correlation.  
For all coupling strengths we have examined, 
$\left<\sigma \cdot \tilde{\sigma}\right>_{\text{ss}}$ 
can be made arbitrarily close to unity by decreasing the various 
$\epsilon$ parameters.  
This level of control ensures that one can generate trial trajectories 
which are correlated with a reference for all times, an essential 
capability for efficient path sampling of long trajectories.

\subsection{WCA Dynamics}

\begin{figure}[ht]
\centering
\includegraphics[width=0.46\textwidth]{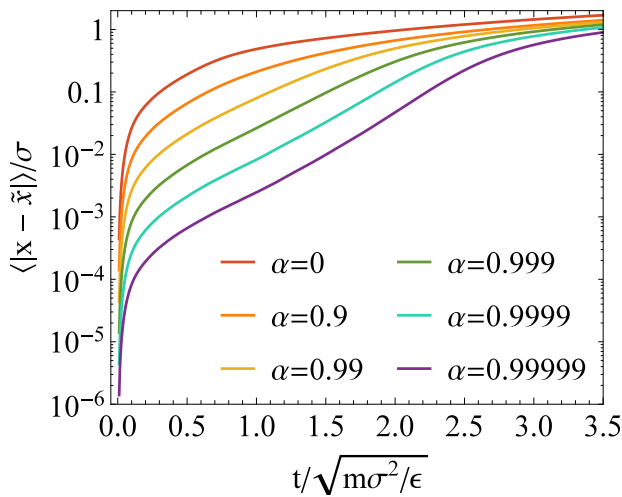}
\caption[Decorrelation of Noise-Guided WCA Trajectories] 
{
  Divergence between reference and trial trajectories of a 
  two-dimensional WCA fluid with Brownian tube noise guidance 
  of strength $\alpha$.
  Average distance between the two trajectories, as defined in 
  Eq.~\ref{eq:wcadist}, is shown for a system of 400 particles with 
  mass $m$ and diameter $\sigma$, in a square box with side length 
  24 $\sigma$ (i.e., density $\rho \sigma^2 = 0.694$). 
  Underdamped Langevin dynamics was propagated with inverse 
  temperature $\beta = 0.2$ and friction coefficient $\gamma = 0.1$
  using a timestep of 0.005 $\tau$, where 
  $\tau = \sqrt{m\sigma^2 / \epsilon}$ and $\epsilon$ is the
  Lennard-Jones interaction energy scale~\cite{Sivak2014}.
  Data are averaged over 500 independent trial trajectories.
}
\label{fig:wca}
\end{figure}

The success we have achieved in synchronizing Ising dynamics with noise 
guidance should not be expected for complex dynamical systems in general.
We demonstrate this limitation for the specific case of a two-dimensional 
WCA fluid~\cite{Weeks1971} evolving by underdamped Langevin dynamics.  
The purely repulsive particles are propagated using an integration scheme 
that requires generating a collection of Gaussian random 
variables~\cite{Athenes2008,Sivak2013}.  
These noises are guided by a Brownian tube proposal, 
Eq.~\ref{eq:stoltzalpha}.  
The similarity between trial and reference noise histories is controlled
by a parameter $\alpha$ that ranges from zero (no noise guidance) 
to one (complete noise guidance).

Starting from identical initial configurations, we propagate dynamics 
with correlated noise histories and monitor the difference between trial 
and reference as
\begin{equation}
\left<\left|\mathbf{x} - \tilde{\mathbf{x}}\right|\right> = \frac{1}{N} \sum_{i=1}^N \left|\mathbf{x}^{(i)} -  \tilde{\mathbf{x}}^{(i)}\right|,
\label{eq:wcadist}
\end{equation}
where $\mathbf{x}^{(i)}$ and $\tilde{\mathbf{x}}^{(i)}$ are the
positions of particle $i$ in the reference and trial and
$\left|\cdot\right|$ is the two-dimensional Euclidean distance.  
At short times, the difference between trial and reference trajectories
is small, but this difference grows exponentially, a hallmark of
chaotic dynamics.  
Even with exceptionally strong noise guidance, trajectories cannot be 
held arbitrarily close to each other for long times, as shown in 
Fig.~\ref{fig:wca}.

Why are we unable to guide the evolution of WCA particles as
effectively as we guided Ising dynamics?  
A principal difference between the two systems is the likelihood of 
spontaneous local recurrences.  
In either case, trajectories with similar initial conditions 
but different noise histories wander away from one another in a global
sense, eventually exploring very different regions of configuration
space. 
Correlating their noise histories generally acts to defer,
but not defeat, this divergence. 
In the case of Ising dynamics, however, a given small block of 
adjacent spins will occasionally align spontaneously in two 
trajectories with reasonable probability. 
Following such a spontaneous, local recurrence, correlated noises again 
work to hold trajectories close.
Global spontaneous recurrence of a large number of spins is, of
course, highly improbable, but the noise guidance seems to ``lock in''
local correlations every time they spontaneously reoccur.

\section{Conclusion}

Transition path sampling has proven useful for a variety of equilibrium, 
as well as non-equilibrium, problems in chemical dynamics.  
The problem of sampling long trajectories, particularly those with
multiple intermediates, has hindered a variety of extensions and 
applications of the methodology.  
We have outlined a modern physical perspective from which to 
assess and address these challenges.
We have demonstrated and discussed successful trajectory guidance in 
the case of Monte Carlo dynamics of an Ising model.
Substantial difficulties remain for systems with continuous degrees of
freedom.

Our results suggest that effective noise guidance of long trajectories
requires a nonnegligible probability of spontaneous local recurrence,
i.e., a significant chance that reference and trial trajectories
transiently align within small regions of space.  
Such synchronization could be particularly helpful for sampling 
reactive trajectories that traverse metastable intermediate states, 
for example the coarsening or assembly of colloidal systems as they 
organize on progressively larger scales.  
In such cases, trial trajectories in the course of path sampling should 
maintain correlations with the reference while passing through the 
intermediates, not just at the endpoints.  
Even without identifying metastable configurations, correlated noises 
could be applied during some intervals but not others. 
A tendency to synchronize would enable trial trajectories to explore 
widely during unguided periods, but to be reined in globally by intermittent
guidance.

We anticipate that these noise-guidance methods will be effective for 
other lattice systems as well, but their usefulness could depend 
sensitively on the exact manner in which the noise influences
dynamics.  
In particular, without incorporating directionality into proposed 
spin changes, we were not able to guide long Ising trajectories.
Furthermore, Ising dynamics can exhibit spontaneous recurrence, 
i.e., transient local alignment between two 
trajectories regardless of their noise histories.  
Because small blocks of Ising spins can adopt only a modest number of 
configurations, such random local synchronization occurs with an 
appreciable probability. 
The probability of recurrence will likely be lower for models 
with a larger collection of possible local configurations, 
e.g., a Potts model or an Ising model with more neighbors.  
We thus expect that the application of noise-guided path sampling could 
face substantial challenges for long trajectories of these more 
intricate lattice models.

\section*{Acknowledgments}
We acknowledge many useful discussions with Michael Gr\"{u}nwald about
strategies for sampling long trajectories.  
T.R.G. acknowledges support from the NSF Graduate Research Fellowship 
and the Fannie and John Hertz Foundation.
P.L.G. was supported by the U.S.  Department of Energy, 
Office of Basic Energy Sciences, through the Chemical
Sciences Division (CSD) of the Lawrence Berkeley National Laboratory
(LBNL), under Contract DE-AC02-05CH11231.

\appendix
\section{Entropy Production Statistics for a One-Dimensional Random Walker with Guiding Forces}
\label{app:guidingforces}
\subsection{Mean Entropy Production}

Here we analytically characterize the entropy production distribution,
$P(\omega)$, for shooting moves generated with guiding forces as 
discussed in Section~\ref{sec:guidingforces}.  
It is useful to first rewrite Eq.~\eqref{eq:entropyproduction} in terms 
of the noise variables,
$\xi$ and $\tilde{\xi}$.  
For a one-dimensional random walk, the position $x_{t+1}$ and the 
difference between reference and trial trajectory, 
$x_{t+1} - \tilde{x}_{t+1}$, can be compactly expressed in terms of 
the noises.
\begin{align}
x_{t+1} &= \sum_{u=0}^t \xi_u\\
x_{t+1} - \tilde{x}_{t+1} &= \sum_{u=0}^t (1-k)^{t-u}(\xi_u-\tilde{\xi}_u).
\label{eq:runningdiff}
\end{align}
After straightforward algebra it is possible to express $\omega$ as
\begin{equation}
\omega = \frac{1}{\sigma^2} \sum_{t=0}^{t_{\text{obs}}-1} \left(S_t^2 - S_t \xi_t^+ \right),
\label{eq:entropyproductionxi}
\end{equation}
where for convenience we have defined 
$\xi_t^+ \equiv \tilde{\xi}_t + \xi_t + 2\mu$, $\xi_t^- \equiv \tilde{\xi}_t - \xi_t$, and
\begin{equation}
S_t \equiv \sum_{u=0}^{t-1} k(1-k)^{t-1-u}\xi_u^-.
\label{eq:Sdef}
\end{equation}
The main text presents results for a random walk without drift, i.e.,
with $\xi$ drawn from a distribution with mean zero. 
Here we consider the more general case with nonzero mean $\mu$.  
Noting that $\left<\xi_t^+\xi_u^+\right> = \left<\xi_t^-\xi_u^-\right> = 2\sigma^2 \delta_{tu}$ 
and $\left<\xi_t^+ \xi_u^-\right> = 0$, the average
entropy production is found to be
\begin{equation}
\left<\omega\right> = 2 \sum_{t=0}^{t_{\text{obs}}-1} \sum_{u=0}^{t-1} k^2 (1-k)^{2(t-1-u)}.
\label{eq:meandoublesum}
\end{equation}
The two geometric series are summed to yield
\begin{equation}
\left<\omega\right> = \frac{2}{(k-2)^2}\left[\left(2-k\right)k t_{\text{obs}}-1 + (k-1)^{2t_{\text{obs}}}\right].
\label{eq:meanscalingapp}
\end{equation}
When $k>2$, $\left<\omega\right>$ grows exponentially in $t_{\text{obs}}$.
This superlinear scaling results from coupling between trajectories 
so strong that the trial trajectory rapidly tends to infinity due 
to a numerical instability, much like the instability that arises 
in conventional molecular dynamics simulations performed with an 
excessively large integration timestep.  
For the useful range of coupling strength, $k < 2$, $\left<\omega\right>
\propto t_{\text{obs}}$ in the long time limit.  
The marginal $k=2$ case is well-behaved 
($\left<\omega\right> = 4 t_{\text{obs}} \left(t_\text{obs}-1\right)$), 
but uninteresting for our purposes.

\subsection{Cumulant Generating Function}
The behavior of the higher-order cumulants can be extracted from the 
cumulant generating function $\ln \left<e^{-\lambda \omega}\right>$. 
This average requires integration over all of the Gaussian $\xi$ and 
$\tilde{\xi}$ variables at all times, which can be performed inductively.  
We define $\phi(\lambda,f, g, h)$ as
\begin{widetext}
\begin{align}
\nonumber
\phi(\lambda, f, g, h, t) &= f \left(\frac{1}{2\sigma \sqrt{\pi}}\right)^{2 t} \int d\xi_0^+ \hdots d\xi_{t-1}^+ d\xi_0^-\hdots d\xi_{t-1}^- \ \exp\left[\frac{1}{\sigma^2} \sum_{i=0}^{t-2}\left(\frac{-(\xi_i^+)^2 - (\xi_i^-)^2}{4} + \lambda S_i \xi_i^+ - \lambda S_i^2\right)\right]\\
& \exp\left[\frac{1}{\sigma^2}\left(\frac{-(\xi_{t-1}^+)^2-(\xi_{t-1}^-)^2}{4} + \lambda \left(S_{t-1} \xi_{t-1}^+ - h S_{t-1}^2 + 2 g k (1-k) S_{t-1} \xi_{t-1}^- + g k^2 (\xi_{t-1}^-)^2\right)\right)\right]
\label{eq:phieq}
\end{align}
\end{widetext}
The integral $\phi$ is defined such that $\left<e^{-\lambda
  \omega}\right> = \phi(\lambda, 1, 0, 1, t_{\text{obs}})$.  
By introducing $f, g,$ and $h$ we can derive recursion relations as we
sequentially integrate out Gaussian noises at the latest remaining
timestep.
In particular, integration over $\xi_{t-1}^-$ then $\xi_{t-1}^+$ returns 
an integral of the same form.  
That is to say 
$\phi(\lambda, f_i, g_i, h_i, t) = \phi(\lambda, f_{i+1}, g_{i+1}, h_{i+1}, t-1)$ 
with
\begin{align}
f_{i+1} &= \frac{f_i}{\sqrt{1-4 \lambda g_i k^2}}\\
g_{i+1} &= \lambda - h_i + \frac{4\lambda g_i^2(1-k)^2 k^2}{1-4 \lambda g_1 k^2}\\
h_{i+1} &= 1 - (1-k)^2 \left((\lambda - h_i) + \frac{4 \lambda g_i^2 (1-k)^2 k^2}{1-4\lambda g_i k^2}\right)
\end{align}
Iterating the map $t_{\text{obs}}$ times corresponds to integrating over 
all of the $2 t_{\text{obs}}$ integrals in Eq.~\eqref{eq:phieq}.
After some algebraic simplification, 
\begin{equation}
\ln \left<e^{-\lambda \omega}\right> = -\frac{1}{2} \sum_{i=0}^{t_{\text{obs}}-1} \ln (1-4\lambda g_i k^2),
\end{equation}
where $g_0 = 0$ and
\begin{equation}
g_{i+1} = \lambda - 1 + \frac{(1-k)^2 g_i}{1 - 4\lambda g_i k^2}.
\label{eq:map}
\end{equation}
The scaled cumulant generating function is then given by
\begin{equation}
\lim_{t_{\text{obs}} \rightarrow \infty} \frac{1}{t_{\text{obs}}} \ln \left<e^{-\lambda \omega}\right> = -\frac{1}{2} \ln(1 - 4\lambda g^* k^2),
\end{equation}
where $g^*$ is a fixed point of the map given in Eq.~\eqref{eq:map}.
Specifically it is the lesser of the two roots of the quadratic
equation obtained when $g_i=g_{i+1}\equiv g^*$ is inserted into
Eq.~\eqref{eq:map}.  
The numerical Legendre transform of this scaled cumulant generating 
function gives the solid black curve in Fig.~\ref{fig:ldffig}(a), 
toward which the results of numerical  sampling should converge for 
long $t_{\text{obs}}$.

\section{Ising Model Steady State Correlations}
\label{app:overlap}

To derive the steady state correlation between reference and trial
trajectories, we examine the time evolution of the probability 
$p(\tau)$ that reference and trial overlap at site $i$ and MC step $\tau$. 
Without loss of generality, we focus on a particular site, $i = 1$.  
Push up/push down moves can be grouped into four classes: 
(i) the reference and trial each act on spin $1$, (ii) the reference
acts on spin $1$ while the trial acts on a different spin, (iii) the
trial acts on spin $1$ while the reference acts on a different spin,
or (iv) neither reference nor trial acts on spin $1$.  
For each case, we derive a transition matrix which maps the 
vector $\left(p(\tau), 1 - p(\tau)\right)$ to its state at MC step 
$\tau + 1$.  The full transition matrix for a step of dynamics is the 
sum of these transition matrices, weighted by the probability of each case,
\begin{align}
\nonumber
T &= \left(1 - \frac{1}{N}\left(1 + \epsilon_{\text{site}} - \frac{\epsilon_{\text{site}}}{N}\right)\right)I + \frac{1}{N} \left(\epsilon_{\text{site}} - \frac{\epsilon_{\text{site}}}{N}\right) Q\\
& \ \ \ \ \ \ \ \ \ \ + \frac{1}{N}\left(\epsilon_{\text{site}} - \frac{\epsilon_{\text{site}}}{N}\right) R + \frac{1}{N}\left(1-\epsilon_{\text{s}} + \frac{\epsilon_{\text{site}}}{N}\right) S.
\label{eq:fulltmatrix}
\end{align}
The transition matrix for case (iv) is the identity matrix, since this 
case cannot alter the overlap at site 1.
The transition matrices for cases (i), (ii), and (iii) are $Q, R,$ and 
$S$, respectively, the forms of which we now derive.

When reference and trial act on different spins, only one copy of 
spin 1 (the reference or the trial) can change its state.  
When the two copies differ at site $i$ after $\tau$ steps, overlap is 
induced with probability 1/4 (i.e., the probability that any given move 
results in a change of spin state). 
For initially aligned copies, loss of overlap similarly occurs with 
probability 1/4. 
This logic applies equally well to cases (ii) and (iii), so
\begin{equation}
Q = R = \begin{pmatrix}3/4 & 1/4\\ 1/4 & 3/4 \end{pmatrix}.
\label{eq:QRdef}
\end{equation}

When both reference and trial act on site 1, we must account for
correlated influence on the two copies.  
As a result, $S$ depends on $\epsilon_{\text{acc}}$ and 
$\epsilon_{\text{dir}}$.  
To enumerate these correlated changes, we denote states
of spin 1 at step $\tau$ in the reference and trial as 
$\sigma_1$ and $\tilde{\sigma}_1$, respectively.  
After the MC step, these spins are given by $\sigma_1'$ and 
$\tilde{\sigma}_1'$.  
Table~\ref{tab:enumeration} lists the possible transformations which 
result in overlapping spins ($\sigma_1' = \tilde{\sigma}_i'$) after 
$\tau+1$ steps.
\renewcommand{\arraystretch}{1.2}
\begin{table}[ht]
\begin{center}
\begin{tabular}{|cccc|cccc|c|}
\hline
$\sigma$ & $d$ & $a$ & $\sigma'$ & $\tilde{\sigma}$ & $\tilde{d}$ & $\tilde{a}$ & $\tilde{\sigma}'$ & Probability\\
\hline
1 & 1 & * & 1 & 1 & 1 & * & 1 & $\frac{p}{2}\left(1-\frac{\epsilon_{\text{dir}}}{2}\right)$\\
1 & 1 & * & 1 & 1 & -1 & 0 & 1 & $\frac{\epsilon_{\text{dir}} p}{8}$\\
1 & 1 & * & 1 & -1 & 1 & 1 & 1 & $\frac{1-p}{4} \left(1-\frac{\epsilon_{\text{dir}}}{2}\right)$\\
1 & -1 & 0 & 1 & 1 & 1 & * & 1 & $\frac{\epsilon_{\text{dir}} p}{8}$\\
1 & -1 & 0 & 1 & 1 & -1 & 0 & 1 & $\frac{p}{4}\left(1-\frac{\epsilon_{\text{dir}}}{2}\right)\left(1-\frac{\epsilon_{\text{acc}}}{2}\right)$\\
1 & -1 & 0 & 1 & -1 & 1 & 1 & 1 & $\frac{1-p}{4}\left(\frac{\epsilon_{\text{dir}}}{2}\right)\left(\frac{\epsilon_{\text{acc}}}{2}\right)$\\
1 & -1 & 1 & -1 & 1 & -1 & 1 & -1 & $\frac{p}{4}\left(1-\frac{\epsilon_{\text{dir}}}{2}\right)\left(1-\frac{\epsilon_{\text{acc}}}{2}\right)$\\
1 & -1 & 1 & -1 & -1 & 1 & 0 & -1 & $\frac{1-p}{4}\left(\frac{\epsilon_{\text{dir}}}{2}\right)\left(\frac{\epsilon_{\text{acc}}}{2}\right)$\\
1 & -1 & 1 & -1 & -1 & -1 & * & -1 & $\frac{1-p}{4}\left(1-\frac{\epsilon_{\text{dir}}}{2}\right)$\\
\hline
\end{tabular}
\end{center}
\caption[Enumeration of Correlated Transitions for Ising Dynamics]
{
Enumeration of moves yielding $\sigma' = \tilde{\sigma}'$.  
Without loss of generality, we only list the moves which start with 
$\sigma = 1$.
The moves starting with $\sigma = -1$ are analogous.
$d$ is the direction of a push with $1$ meaning up.  
$a$ indicates whether the move is accepted (1) or rejected (0). 
An asterisk indicates that both options yield the same result.
}
\label{tab:enumeration}
\end{table}
Collecting terms in the table and making use of the fact that $S$ 
is a probability-conserving transition matrix, we find
\begin{equation}
S = \frac{1}{2} \begin{pmatrix} 1 + \left(1 - \frac{\epsilon_{\text{dir}}}{2}\right)\left(1 - \frac{\epsilon_{\text{acc}}}{2}\right) & \ \ 1 - \frac{\epsilon_{\text{dir}}}{2}\left(1 - \frac{\epsilon_{\text{acc}}}{2}\right)\\
1 - \left(1 - \frac{\epsilon_{\text{dir}}}{2}\right)\left(1 - \frac{\epsilon_{\text{acc}}}{2}\right) & \ \ 1 + \frac{\epsilon_{\text{dir}}}{2}\left(1 - \frac{\epsilon_{\text{acc}}}{2}\right)
\end{pmatrix}.
\label{eq:Smatrix}
\end{equation}

Propagation according to the transition matrix $T$ gives the overlap
probability after a single MC step:
\begin{equation}
\begin{pmatrix}
p(\tau+1)\\1-p(\tau+1)
\end{pmatrix} = T \begin{pmatrix}
p(\tau)\\1-p(\tau) \end{pmatrix}
\end{equation}
The first row of this matrix equation reads, after some algebra, 
\begin{widetext}
\begin{equation}
p(\tau+1) = p(\tau) - \frac{p(\tau)}{N} + \frac{\epsilon_{\text{site}} - \frac{\epsilon_{\text{site}}}{N}}{2N} + \frac{1-\epsilon_{\text{site}}+\frac{\epsilon_{\text{site}}}{N}}{2N} \left[1 - \frac{\epsilon_{\text{dir}}}{2} \left(1 - \frac{\epsilon_{\text{acc}}}{2}\right) + \left(1 - \frac{\epsilon_{\text{acc}}}{2}\right) p(\tau)\right].
\label{eq:overlapeom}
\end{equation}
\end{widetext}
We are interested in the steady state solution, found by setting 
$p(\tau) = p(\tau+1)$.  
Multiplying the equation through by $N$, followed by algebraic 
simplification, yields Eq.~\ref{eq:hightemp} of the main text.


%

\end{document}